\documentclass[epsfig,11pt]{article}
\usepackage{amsmath,amssymb}
\usepackage{epsfig}
\usepackage{graphicx}
\usepackage{array}
\usepackage{color}
\usepackage{bm}
\usepackage{bbm}

\usepackage{threeparttable}
\usepackage{adjustbox}
\usepackage[nottoc,notlof,notlot]{tocbibind} 
\usepackage[titles]{tocloft} 
\usepackage{booktabs}
\usepackage[justification=centering]{caption}
\usepackage[T1]{fontenc}
\usepackage{makeidx} 
\makeindex

\newcommand{\beq}{\begin{equation}}   
\newcommand{\eeq}{\end{equation}}
\newcommand{\beqn}{\begin{eqnarray}}   
\newcommand{\eeqn}{\end{eqnarray}}

\newcommand{\gsim}{\lower.7ex\hbox{$
\;\stackrel{\textstyle>}{\sim}\;$}}
\newcommand{\lsim}{\lower.7ex\hbox{$
\;\stackrel{\textstyle<}{\sim}\;$}}
\begin{document}

\newpage

\begin{center}

{\bf\large  FOREWORD TO THE SECOND EDITION OF\\[1mm]
 ``THE SUPERSYMMETRIC WORLD''\,$\biguplus$
}

\vspace{2mm}

{\large M.  SHIFMAN}

\vspace{2mm}

{\small \em Theoretical Physics Institute, University of Minnesota,
Minneapolis, MN 55455}

{\tt shifman@umn.edu} 
\end{center}

The First Edition of this book was released in 2000, just before the symposium ``Thirty Years of Supersymmetry''  was held at the William I. Fine Theoretical Physics Institute (FTPI) of the University of Minnesota. Founders and trailblazers of supersymmetry descended on FTPI, as well as a large crowd of younger theorists deeply involved in research in this area. Remarkably, it was at this  event that many of the early pioneers of the field met face-to-face for the first 
time.\! Table\! \ref{table:1.1}  presents the Contents of the historical part of the SUSY-30 Proceedings   \cite{30Proc}. 
Since then 23 years have elapsed and significant  changes happened in supersymmetry (SUSY). Its history definitely needs an update. 

Below the reader will see a table which can be viewed as a starting point for the current edition.
 In 2000 supersymmetry explorations were on the rise and
SUSY-based phenomenology attracted hundreds of young researchers, with around 2600 original papers published annually 
(see the graphs in Figs. \ref{fig:pgs3},\,\ref{fig:pgs4} \!on page
\pageref{fig:pgs4}).\!\! Expectations of an imminent experimental discovery of superpartners were prevailing in the community; the general mood was rosy.

\begin{table}[htbp!]
	\normalsize
	\centering
	\caption{\small The historic part in   \cite{30Proc}. Table of contents}
    \label{tab:table1}
	\begin{adjustbox}{max width=\textwidth}
		\begin{tabular}{ll}
		\toprule
		\textbf{Part 1: Supersymmetry from the East}          &  		      \\
		\midrule
		{\sl N. Koretz-Golfand }                         & Supersymmetry -- 30	    \\
		  {\sl E.P. Likhtman}                              & 	Around SUSY 1970				     					\\
		         {\sl       V. Akulov}                 & Nonlinear way   SUSY and ${\cal N}$ extended SUSY			\\ 				
	{\sl S.I. Volkova, A.A. Zheltukhin }                     & Glimpses of Dmitry Volkov life and work				    						\\
		         {\sl V.A. Soroka}                       & Supersymmetry and the odd Poisson bracket								\\
		         \midrule
		   \textbf{Part 2: Supersymmetry from the West}          &  		      \\
		    \midrule
		      {\sl Pierre Ramond}                       & Boson-fermion confusion: The string path to SUSY	\\
		       {\sl J.H. Schwarz}                       &String theory origins of supersymmetry \\
		       {\sl J.-L. Gervais}                       & Symmetries: early days and nowadays\\
		       {\sl B. Sakita} & Symmetries of fermions in the lowest Landau level\\
		       {\sl P. Fayet} & About the origins of supersymmetric standard model\\
		       {\sl J. Iliopoulos} & Non-renormalization theorems in global supersymmetry\\
		       {\sl L. O'Raifeartaigh} & {Chiral fermions on the lattice}\\
		       {\sl P. West} & Non-renormalization theorems in supersymmetric theories\\
		       {\sl M.F. Sonius} & Recollections of a young contributor\\
		       {\sl S. Ferrara} & Superconformal algebras and supergravities in higher dimension\\
		       {\sl A.H. Chamseddine, R. Arnowitt, P. Nath}& Supergravity unification\\
		       {\sl Bruce de Witt}& Elrctric-magnetic dualities in supergravity\\
		\bottomrule
		\end{tabular}
	\end{adjustbox}
	\label{table:1.1}
\end{table}

\newpage 

\begin{figure}[h]
\begin{center}
\includegraphics[width=10cm]{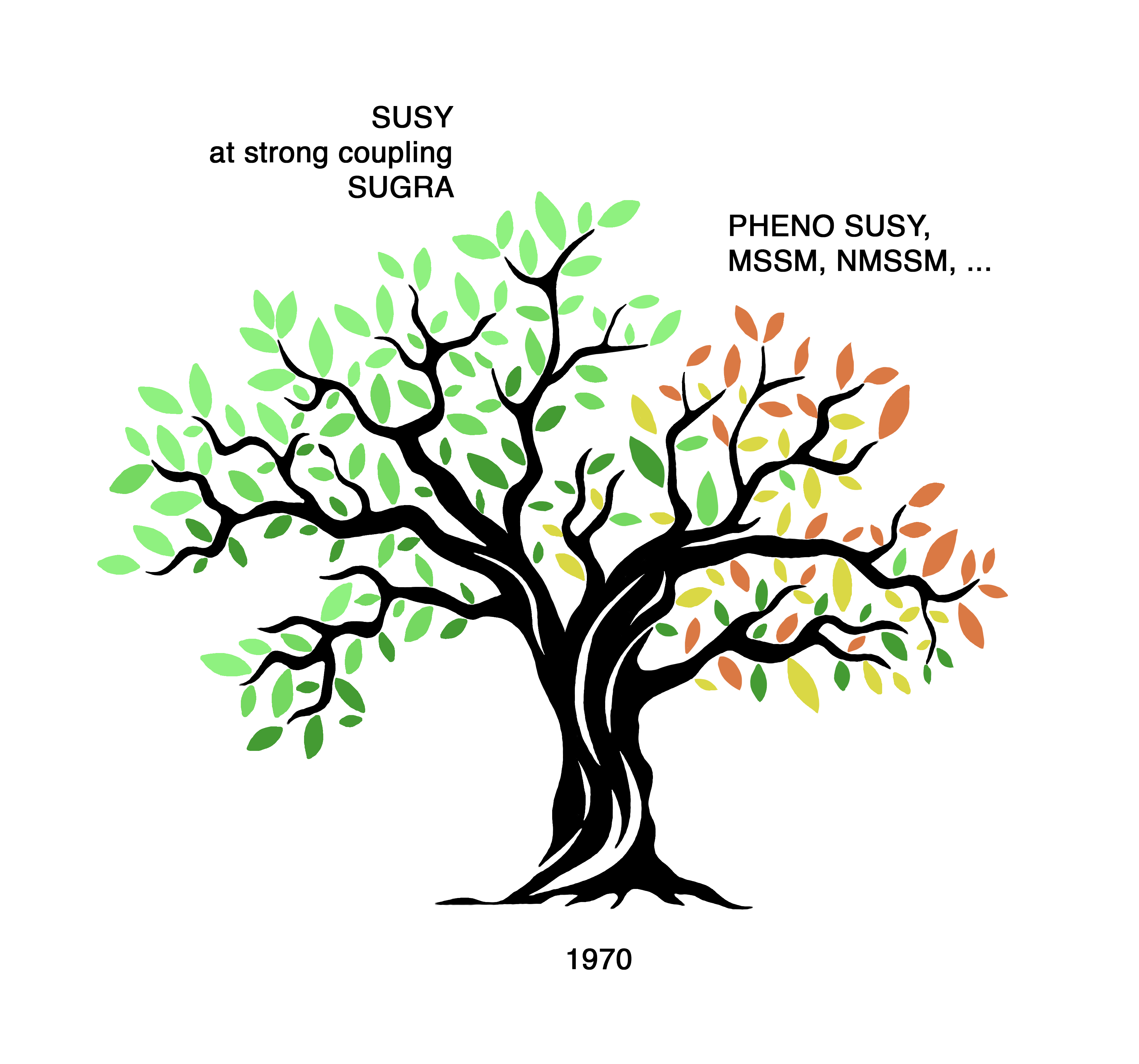}
\vspace{-10mm}
\end{center}
\caption{\small
The superworld tree in 2023. {\scriptsize{\copyright M. Shifman}}}
\label{001}
\end{figure}
\label{007}

The modern perspective on supersymmetry looks different. So far, all searches of SUSY in LHC experiments have brought negative results, a great disappointment.
The rosy mood of the early 2010s faded away.

On the other hand, the unique power of supersymmetry became obvious in mathematical physics, string mathematics, and such topics as black holes and critical (BPS) solitons, especially at strong coupling where SUSY hardly has any competitors. For the first time ever, the analytic proof of quark confinement was found 
in ${\cal N}=2$ super-Yang-Mills theory   \cite{Seiberg:1994rs} slightly deformed by an ${\cal N}=1$ perturbation. 

This is reflected in Fig.~$\!$\ref{001} which I designed especially for the current edi\-tion. (See also the graph in Fig. \ref{fig:pgs5} on page \pageref{fig:pgs5}). 
 Figure \ref{001} demonstrates that the ``Pheno'' branch of supersymmetry is drying out -- its future is unclear at the moment. 
The ``strong coupling'' branch  dealing with ``imaginary''  worlds and more mathematical applications of SUSY continues to grow.

Perhaps, it would be fair to add that the past searches for superpartners and the studies of very feebly-interacting goldstino/gravitino in a susy-breaking sector 
led to current searches for new feeble interactions and dark sector. Life goes on, work continues, in particular, the hunt for axion-like particles (ALPs).
One may hope that a new tree will sprout in Fig. \ref{001}.

``Supersymmetry is so beautiful and suggestive that most of us think it has got to show up sometime in nature, although so far it hasn't,'' said Steven 
Weinberg\index{Weinberg, Steven}  in his 1979 Nobel speech. I still believe in Weinbeg's prophecy but it is reasonable to ask ``when?'' Unfortunately, I am not a prophet and the only answer I can give is ``The future's not ours to see...''

\begin{quote}
\small

``We cannot guess the way of word\\
In real world, how it'll return...'' \footnote{A quotation from a Russian poet Fyodor Tyutchev\index{Tyutchev} (1803-1873). Translated by Lyudmila Purgina.}

\vspace{-2mm}

\end{quote}

In this volume we collected personal memories describing days long gone, with their excitement and even euphoria.\footnote{Or was it inebriation?..} These testimonies were penned by the discovers of supersymmetery, trailblazers, and pioneers themselves. Hopefully they will be used by future scholars, in particular, historians of science. Unfortunately, even in 2000 when the First Edition was being prepared it was too late to obtain personal memoirs from Yuri Golfand\index{Golfand} (1922-1994) and 
Abdus Salam\index{Salam} (1926-1996). We had to settle for their students and collaborators -- Evgeny Likhtman\index{Likhtman} in the first case and John Strathdee\index{Strathdee}\label{Strathdee} and Peter West\index{West} in the second.\footnote{Here I have to make a few explanatory remarks. (i) In English literature Golfand's surname is written in two different ways -- Golfand\index{Golfand} or Gol'fand. The apostrophe in the latter version indicates that a certain ``mute'' letter of the Russian alphabet is omitted. Phonetically this omitted letter makes the previous ``l'' soft, as in French;
(ii) Evgeny Likhtman\index{Likhtman} was Golfand's PhD student and a co-author of three crucial papers;  (iii) John Strathdee\index{Strathdee} was 
Salam's\index{Salam} co-author of 27 years! He retired from ICTP in the late 1990s and hid somehwhere in New Zealand. Finding him there was not easy;  I was told that  he had left his mailing address only to a certain secretary at ICTP. I contacted her and eventually she entrusted me with Strathdee's address after I promised her to use it just once. Strathdee's reply to my letter contained seven lines, for which I am grateful, see Chapter 5.}

\vspace{-2mm}

\begin{center}
*****
\end{center}

\vspace{-3mm}

Almost all of the early explorers of supersymmetry, its founding fathers, have left our world after 2000. This is a sad  -- but alas...--  a natural process.
Julius  Wess\index{Wess} and Bruno Zumino\index{Zumino}, who (among others) were the  masterminds and discoverers of four-dimensional supersymmetry, died in  2007 and 2014 respectively. They were  life-long friends and collaborators. I feel that  Julius and Bruno, who gave so much to our community,  deserve a special ``In Memoriam'' chapter in the current edition (see Chapter ~1).

The available literature about the dawn of supersymmetry has significantly expanded since 2000. A wealth of information is presented in   \cite{30Proc} and  the accompanying volume   \cite{YG}. In 2004, a  fascinating ``Concise Encyclopedia of Supersymmetry'' was published   \cite{concise} with more than 700 articles and a historical section written specifically for this Encyclopedia by E. Likhtman,\index{Likhtman}  Dmitry Volkov,\index{Volkov} V. Akulov,\index{Akulov} 
H. Miyazawa,\index{Miyazawa} G. Stavraki,\index{Stavraki} V. Kac,\index{Kac} V. Pakhomov,\index{Pakhomov} J. \L{}opuszanski,\index{Lopuszanski} 
R. Haag,\index{Haag} and D. Leites.\index{Leites} Next, it is worth mentioning the collection   \cite{berezin}  narrating the story of Felix Berezin,\index{Berezin}
 an outstanding mathematician who created a mathematical apparatus used in supersymmetric field theories.
 The integral over the anticommuting Grassmann variables that he introduced in the 1960s paved the way for the path integral formulation of quantum field theory with fermions, the heart of modern supersymmetric field theories and superstrings.\footnote{See Marinov's\index{Marinov} article in Chapter 6.} The Berezin integral is named for him, as is the closely related construction of the Berezinian, which may be regarded as the superanalogue of the determinant. In the same year the Dmitry Volkov\index{Volkov} Memorial Volume   \cite{volkov}
was published in Ukraine in Russian. It contains both recollections of his colleagues and friends and Volkov's selected interviews spanning the years 1986-1991.
Two detailed essays on Yuri Golfand\index{Golfand} appeared in Part 2 of the book   \cite{mad}.\label{streamrec} 

In 2012 a profound collection {\sl The Birth of String Theory}   \cite{birthst} was released. It contains a wealth of information on the early days of string theory some of which is relevant to our Chapter 2; see in particular Ramond's\index{Ramond} articles.

Approximately a year before his death Stanley Deser\index{Deser} 
published a book of recollections about his life and work   \cite{SD}. The chapter entitled  ``A Big Year''  will be especially interesting to the reader since it narrates Deser's viewpoint on the story of the creation of Supergravity (SUGRA). Other creators of SUGRA published a number of reviews containing their recollections.
The appropriate excerpts will be presented in Chapter 4 of the current Edition. 

Finally, a remarkable connection between the late works of W. Heisenberg\index{Heisenberg} and the discovery of nonlinearly realized supersymmetry by 
Volkov\index{Volkov} and Akulov\index{Akulov} was revealed  in   \cite{HVS}.

Compared to the 2000 edition, the Second Edition is significantly expanded. First, as was mentioned, we add Chapter 1, {\sl In Memoriam: Julius Wess and Bruno Zumino},\index{Wess}\index{Zumino}
which presents recent recollections on Zumino and Wess. Chapter 4, {\sl Local Supersymmetry (Supergravity)}, is completely revised, and now includes a number of articles
which appeared after 2000. Chapter 7 containing a brief report on SUSY-50\,\footnote{The conference {\sl 50 Years of Supersymmetry} (SUSY 50) was organized 
by FTPI, University of Minnesota, on May 18-20, 2023.}\index{SUSY-50} was prepared specifically for the Second Edition. In the newly organized Appendix the reader can find English translations of some of the early papers by Golfand\index{Golfand}, Likhtman,\index{Likhtman} Volkov,\index{Volkov} Akulov,\index{Akulov} and 
Soroka\index{Soroka} which were published in JETP and therefore are not easily accessible. In a number of instances I added comments or explanatory remarks, see e.g. pp. 248 and 452.

Other additions include 
an essay of John Iliopoulos\index{Iliopoulos} on the early results, dreams and expectations of supersymmetry, as well as a brief review  of the present situation (Chapter 5).
In the same Chapter the reader will find recollections about Viktor Ogievetsky\index{Ogievetsky} (1928-1996), the inventor of the harmonic superspace for ${\cal N}=2$ theories. 
Upon reflection, I decided that supersymmetry at {\em strong coupling} should be viewed as a ``special branch'' of SUSY (Fig. \ref{001}), and a few words about its pioneers are in order in the Second Edition. Correspondingly, I added an interview with Nati Seiberg\index{Seiberg} in Chapter 5 devoted to the pioneers as well as my commentary. 

Each article which was absent in the First Edition is marked by $\biguplus$.

I also correct two mistakes.
The Foreword to the First Edition stated:

\vspace{-2mm}
\begin{quote}
\small 
Often students ask where the name ``supersymmetry'' came from.
It seems that it was coined in the paper by Salam and Strathdee   \cite{SS}\index{Salam}\index{Strathdee} 
where these authors constructed supersymmetric Yang-Mills
theory [using superspace formalism]. This paper was received by the editorial office on June 6, 1974,
exactly eight months after that of Wess and Zumino.\index{Wess}\index{Zumino} Super-symmetry (with a hyphen) is in the title, while in
the body of the paper Salam and Strathdee use both, the old version of Wess
and Zumino, ``super-gauge symmetry,'' and the new one. 
\end{quote}
\vspace{-2mm}
When Bruno Zumino read this paragraph (it was long ago) he got upset and replied me in a letter which also contained
a typewritten text of his talk   \cite{bztalkss}.

He wrote:
\vspace{-2mm}
\begin{quote}
\small
Dear Misha,
\label{page6}

[...] I did the same in my talk at 17th International Conference on High-Energy Physics. Please, read it.
\end{quote} 

\vspace{-2mm}

I apologized and promised Bruno to correct my mistake. Now, many years later I fulfill the promise. The reader can find Bruno's 1974  talk added in 
Chapter 3,  just after Julius Wess' talk.
If the reader looks through the above Bruno's talk he or she will note that in fact, apparently, Zumino and Salam had a pre-publication communication.
The issue of Zumino's priority remains foggy. Nevertheless, the talk is worth reading.

Another unintentional omission was pointed out to me by Pierre Ramond.\index{Ramond} He suggested supplementing the narrative on 
Pierre Fayet's\index{Fayet} important contribution (MSSM),\footnote{Namely, in the  First Edition one reads:
\begin{quote}
In the subsequent years, Pierre Fayet combined the Brout-Englert-Higgs\index{Brout}\index{Englert}\index{Higgs} mechanism with supersymmetry and introduced superpartners to the  Standard Model particles   \cite{PF}. This was the inception and birth of the minimal supersymmetric standard model (MSSM). Heavy quarks unknown in the 1970s,
neutrino mixing angles and some details were introduced somewhat later [...],
\end{quote}}
by the reference to Dimopoulos and Georgi   \cite{DG}\index{Dimopoulos}\index{Georgi} who developed a supersymmetric model with softly broken SUSY  in the context of Grand Unification in 1981.\index{Grand Unification}

\begin{figure}[h]
\begin{center}
\includegraphics[width=7cm]{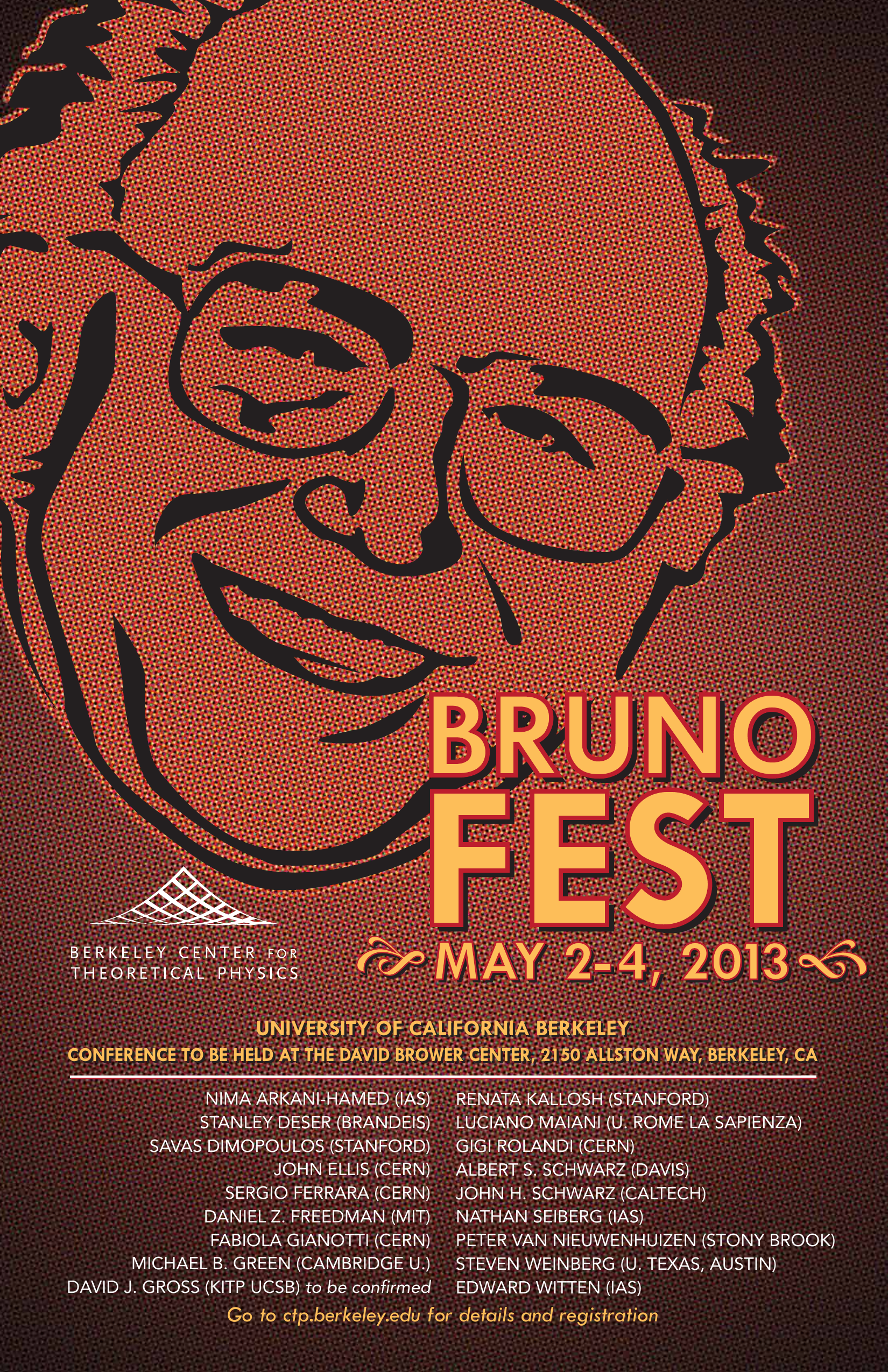}
\end{center}
\caption{
One of the posters honoring Bruno Zumino's 90th anniversary.}
\label{002}
\end{figure}
\label{zposter}

%\newpage

As was mentioned on page \pageref{Strathdee}, Abdus Salam was one of the SUSY pioneers whose role was absolutely crucial. He (and his life-long collaborator John Strathdee) invented superspace and superfields formalism   \cite{SaSt} (this paper was received by the Editorial Office on February 26, 1974). Ferrara-Wess-Zumino work on the same topic   \cite{FWZ} was issued a little later (being received by the Editorial Office on May 3, 1974, it cites   \cite{SaSt}). There are other important overlaps between Ferrara, Wess and Zumino and Salam and Strathdee, e.g. the discovery of super-Yang-Mills theory (see   \cite{FZ}, received by the Editorial Office on May 27, 1974)\footnote{There is a curious typo in the published version of   \cite{FZ}. The editorial note reads ``Received 27 May 1973.'' If correct, the arrival of the manuscript to the Nuclear Physics Editorial Office in May 1973  would be an acausal effect. As far as I know, this typo has never been corrected.},   and   \cite{asalam}, received on June 6, 1974). In the First Edition brief recollections of Abdus Salam were presented only in the 
articles by Martin Sohnius and Peter West. For the Second Edition I solicited a more detailed essay from Peter West, and he kindly agreed (see 
page 249).

I have to add a remark on the 75-pages historical essay (p. 355) written by Rosanne Di Stefano 37 years ago which was published in the First Edition. A stream of relatively recent recollections authored by the Creators and Pioneers of SUSY and string theory  (see p. \pageref{streamrec}) made some parts of this essay dated, especially Sections 8 and 9. Upon reflection, I decided to keep it intact in the current Second Edition because it contains a large number of details and names regarding the early days which are rapidly fading away in community's memory. What is even more important, it gives an idea of the prevailing trends in the community in the 1980s -- quite different from those we witness today almost 40 years later (Chapter 7).

Preparing the Second Edition I noticed a change in the perception of SUSY history which is social rather than scientific. In the 1970, 1980s, and 1990s only a few early
researchers in this area were aware of the contributions of Golfand,\index{Golfand} Likhtman,\index{Likhtman} Volkov,\index{Volkov} Akulov\index{Akulov} and 
Soroka.\index{Soroka} Virtually none of the young theorists knew  about works on four-dimensional  supersymmetry  and supergravity 
``before Wess-Zumino.''\index{Wess}\index{Zumino}
%\marginpar{\tiny\color{red} Make ref. to Ramond.}
This was one of the dark consequences of the 70-year Iron Curtain between the USSR and the rest of the world erected by the communist regime. 
The proceedings of SUSY-30   \cite{30Proc}, the subsequent publications (the corresponding references can be found on page \pageref{glvas}) and a number of talks
on the origins of SUSY I gave in the last two decades on various occasions led to a general recognition of the ``contribution from the East.'' Now it has become common knowledge\,\footnote{See e.g. Chapter 2, Ramond's\index{Ramond} essay on p. 65.} which I view as my modest achievement  in restoring ``fair attribution'' practice in our community.

Concluding this Foreword I have to mention that, unfortunately, Gordy Kane\index{Kane} was unable to participate in preparation of the current edition. Therefore, I am solely responsible for all inaccuracies  the reader may find in the added material.

\section*{Acknowledgments}

With great pleasure I acknowledge  many illuminating discussions with Pierre Fayet,\index{Fayet} Dan Freedman\index{Freedman} and Pierre Ramond. 
I am grateful to Ivo Sachs,\index{Sachs} Pierre Ramond,\index{Ramond}  S. Moskaliuk,\index{Moskaliuk} Hermann Nicolai,\index{Nicolai} 
Mary K. Gaillard,\index{Gaillard}  John Iliopoulos,\index{Iliopoulos} K. S. Stelle,\index{Stelle} Evgeny Ivanov,\index{Ivanov} Emiry Sokatchev,\index{Sokatchev}
Natan Seiberg,\index{Seiberg} Stanley Deser,\index{Deser}  Bayram Tekin,\index{Tekin} Peter van Neuwenhuzen,\index{Nieuwenhuizen, van} Daniel Freedman,\index{Freedman} and S. Duplij\index{Duplij} for their contributions to the current version of this collection.

The cover of this book is based on a fragment from the painting {\sl Victory over Eternity} by Pavel Filonov\index{Filonov} (1883-1941), a Russian artist, one of the leaders of the ``Russian Avant-garde.''\footnote{ Filonov  died of starvation  during the Nazi Siege of Leningrad. Perhaps, I should mention here a curious ``physics story'' regarding Filonov's painting {\sl Shostakovich's Symphony}. An inscription on the canvas read ``19 January... eclipse 15-39 18 total eclipse.'' In 1986, a well-known nuclear physicist Yakov Smorodinsky\index{Smorodinsky} learned with surprise \cite{nsmor} that none of the art experts could determine when exactly Filonov created this famous work.
Smorodinsky looked up the dates and times of the eclipses  which could be observed in Russia during Filonov's life; the lunar eclipse of 1935  fitted the data on the canvas. Thus, he managed to establish the exact date. Since then, Filonov's {\sl Shostakovich's Symphony}  is universally  dated by 1935, with reference to Smorodinsky, see e.g. Irina Pronina, {\em We
cannot guess the way our words will echo through the ages...}, in the Collection {\sl Ya. A.
Smorodinsky, Selected Works}, Classics of Science Series (URSS, Moscow, 2006, ISBN 978-
5-8360-0537-5) page 537, in Russian. I am grateful to A. Kataev who pointed out the above reference to me.} 

 I use this opportunity to say thank you to Vera Kessenich\index{Kessenich} who kindly helped me to obtain the reprint permission from the State Russian Museum in Petersburg.

I am very grateful to Lakshmi Narayanan\index{Narayanan} -- my World Scientific editor for many years -- with whom I worked 
on around a dozen projects. Now, when she is retiring I would like to say thank you, Lakshmi.

\vspace{2mm}

This work is supported in part  by DOE grant DE-SC0011842 and the Simons Foundation Targeted Grant 920184  to the Fine Theoretical Physics Institute.

\newpage

\newpage

\begin{center}

{\bf COMMENT ON CHRONOLOGY AND NUMEROLOGY OF\\[1mm] RESEARCH  ON
SUPERSUMMETRY -- 2023}\,$\biguplus$
\setcounter{figure}{0}

\vspace{2mm}

{M.  SHIFMAN}

\vspace{2mm}

{\small \em Theoretical Physics Institute, University of Minnesota,
Minneapolis, MN 55455}

\vspace{1mm}

{\tt shifman@umn.edu}
\end{center}

I will refer to the ``Prehistoric Era''
all works relevant to supersymmetry which were done before the first 
publication of Wess and Zumino on this subject.

\vspace{0.1mm}

These papers are:

\vspace{0.2mm}

\noindent
(1) {\bf Yu.A. Golfand and E.P. Likhtman}, {\it
Extension  of the Algebra of Poincar\'{e} Group Operators and Violation
of  P-Invariance}, {JETP  Lett.} {\bf 13},   323--326 (1971).
Received by the Editorial Office  on March 10, 1971;\\[1mm]
(2)
{\bf Yu.A. Golfand and E.P. Likhtman}, 
{\it On the Extensions of the Algebra of the Generators of the
Poincar\'e Group by the Bispinor Generators}, in 
I. E. Tamm Memorial Volume
{\em Problems of Theoretical Physics}, Eds. V.L. Ginzburg {\em et al.},
(Nauka, Moscow 1972), pp. 37--44.\footnote{English translation is published in {\sl The Many Faces of the Superworld}, Ed. M. Shifman (World Scientific, Singapore, 2000), p. 45.} ``Signed (authorized) for print'' on 
March 20, 1972;\\[1mm]
(3) {\bf E.P. Likhtman}, 
{\it Irreducible Representations
of the Extension of the Algebra of the Poincar\'e Group Generators by 
the Bispinor Generators},
 Report of the Lebedev Physics Institute \# 41, 1971,
pp. 1--15.\footnote{English translation of this preprint is published in {\sl SUSY-30}, Proc. Int. Symposium Celebrating 30 Years of Supersymmetry, Eds. K. Olive, S. Rudas, M. Shifman, Nucl. Phys. B (Proc. Suppl.) 101 (2001), p. 9.} ``Signed (authorized) for print''
on  April 12, 1971;\\[1mm]
(4) {\bf A. Neveu and  J.H. Schwarz, }
{\it Factorizable Dual Model of Pions}, 
 Nucl. Phys. {\bf B31}, 86--112 (1971). 
Received by the Editorial Office  on March 25, 1971;\\[1mm]
(5) {\bf P. Ramond},
{\it Dual Theory for Free Fermions}, 
Phys. Rev. {\bf D3}, 2415--2418 (1971).
Received by the Editorial Office on January 4, 1971;\\[1mm]
(6)
{\bf  A. Neveu, J.H. Schwarz, and C.B. Thorn,}
{\it Reformulation of the Dual Pion Model},
 Phys. Lett. {\bf 35B},  529--533 (1971). 
Received by the Editorial Office on May 7, 1971;\\[1mm]
(7) {\bf J.-L. Gervais and B. Sakita,}
{\it Field Theory Interpretation of Supergauges in Dual Models},
 Nucl. Phys. {\bf B34}, 632--639 (1971).
Received by the Editorial Office  on August 13, 1971;\\[1mm]
(8)
 {\bf D.V. Volkov and  V.P. Akulov, }
{\it Possible Universal Neutrino Interaction},  JETP
Lett. {\bf 16}, 438--440 (1972). 
Received by the Editorial Office  on October 13, 1972; \\[1mm]
(9)
{\bf D.V. Volkov and V.P. Akulov,}
{\it Is the Neutrino a Goldstone Particle?}
 Phys. Lett. {\bf B46}, 109--110 (1973).
Received by the Editorial Office  on March 5, 1973; 
(10)
{\bf  D.V. Volkov and V.A. Soroka,}
{\it Higgs Effect for Goldstone Particles with Spin
1/2}, JETP Lett. {\bf 18},
312--314 (1973).
 Received by the Editorial Office  on September 3, 1973;\\
(11) {\bf J. Wess and  B. Zumino,}
{\it  Supergauge Transformations in Four Dimensions},
 Nucl. Phys. {\bf B70}, 39--50 (1974).
Received by the Editorial Office  on October 5, 1973.
\begin{figure}[t]
  \begin{center}
  \includegraphics[width=3.8in]{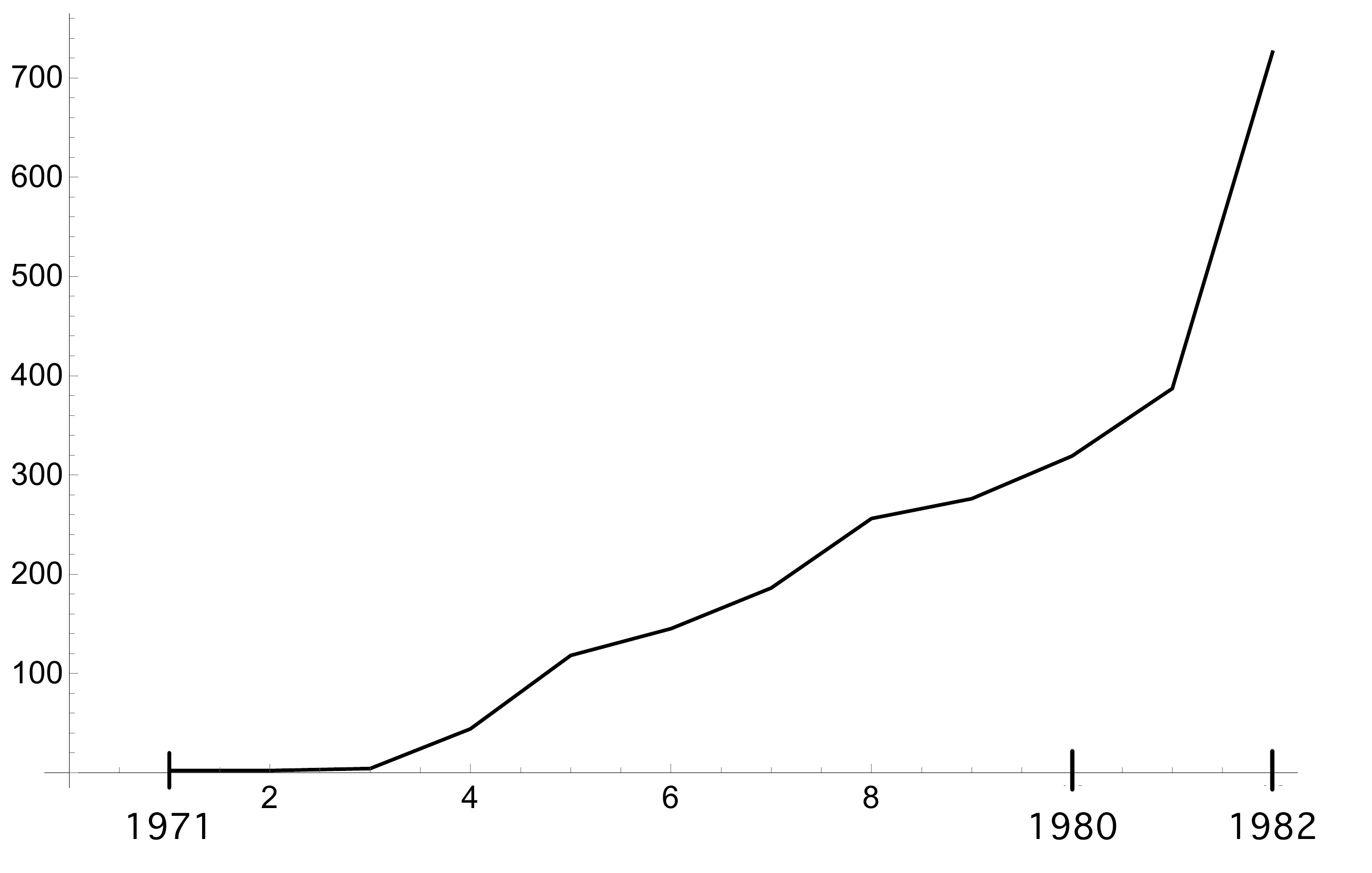}
  \vspace{-3mm}
     \end{center}
  \caption
    {\small The number of papers on supersymmetry published between 1971 and  1982. The beginnings of the accelerated growth
in 1973 and even more accelerated growth in 1982 are visible.}
     \label{fig:pgs_1}
\end{figure}

The dates quoted for the papers (1)-(3)  and (8)-(10)
 must be considered as upper bounds rather than the 
actual submission dates. This is due to the fact that
all materials intended for publication in the USSR had to be pre-cleared by  various
censoring offices, including the omnipresent GLAVLIT, the secret  governement 
agency in charge of the final clearance. A latent period of a few months
was unavoidable. During this time the paper prepared for publication was officially non-existent.\footnote{According to recollections of Likhtman 
(p.\!\! 114) and Marinov (p.\!\! 344) Golfand's superalgebra on p.\!\! 450 was found in 1969.} 
\begin{figure}[t]
  \begin{center}
  \includegraphics[width=3.8in]{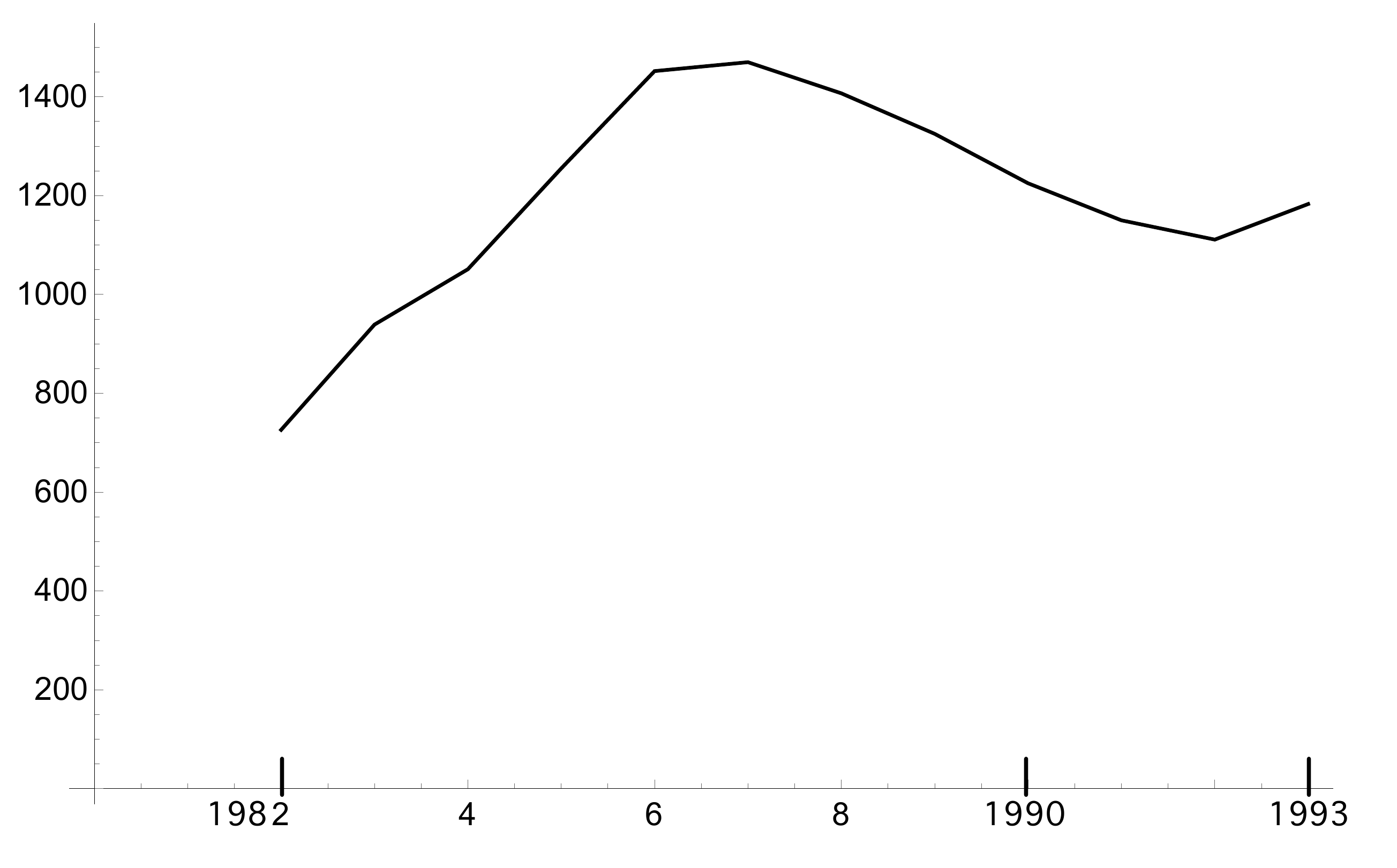}
   \vspace{-5mm}
  \caption
    {\small
The number of papers on supersymmetry published between 1982 and  1993. Note the maximum at the level of 1400 papers in 1987
and a local minimum in 1992-1993.
}
   \end{center}
\label{fig:pgs2}
\end{figure}
In the works (5)-(7) authored by the string-theory pioneers  two-dimensional supersymmetry was introduced on the string world sheet.
To this end two-dimensional fermion  fields were added on the world sheet. They became building blocks for supersymmetries of various types 
in string theory. 
At this point it is worth quoting John Schwarz's  clarification (see Chapter 2),
\begin{quote}
{\small The Ramond-Neveu-Schwarz string theory actually contains local space-time
supersymmetry, but we were slow to realize that.
The realization came after supersymmetry in four dimensions was 
thoroughly studied  by others.}
\end{quote}

The last paper (11) in the above list presents an independent discovery
of supersymmetric field theory in four dimensions.
It opens the ``Historic Era'' of supersymmetry.

\begin{figure}[h]
  \begin{center}
  \includegraphics[width=3.5in]{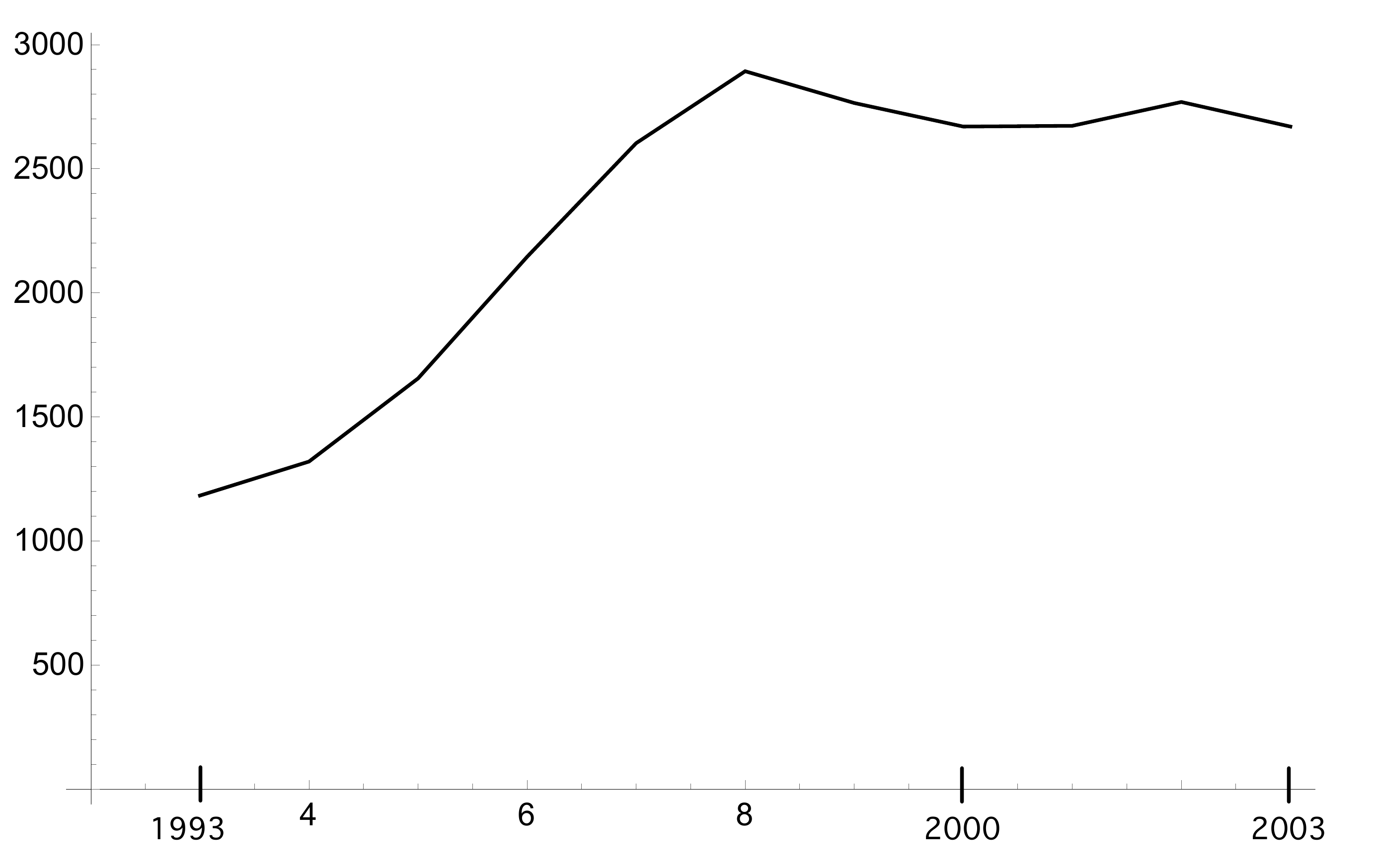}
      \end{center}
         \caption{\footnotesize
The number of papers on supersymmetry published in 1993-2003. The absolute maximum at the level of 2900 papers per year
is achieved in 1998-1999.}
\label{fig:pgs3}
\end{figure}

\begin{figure}[h]
  \begin{center}
  \includegraphics[width=3.5in]{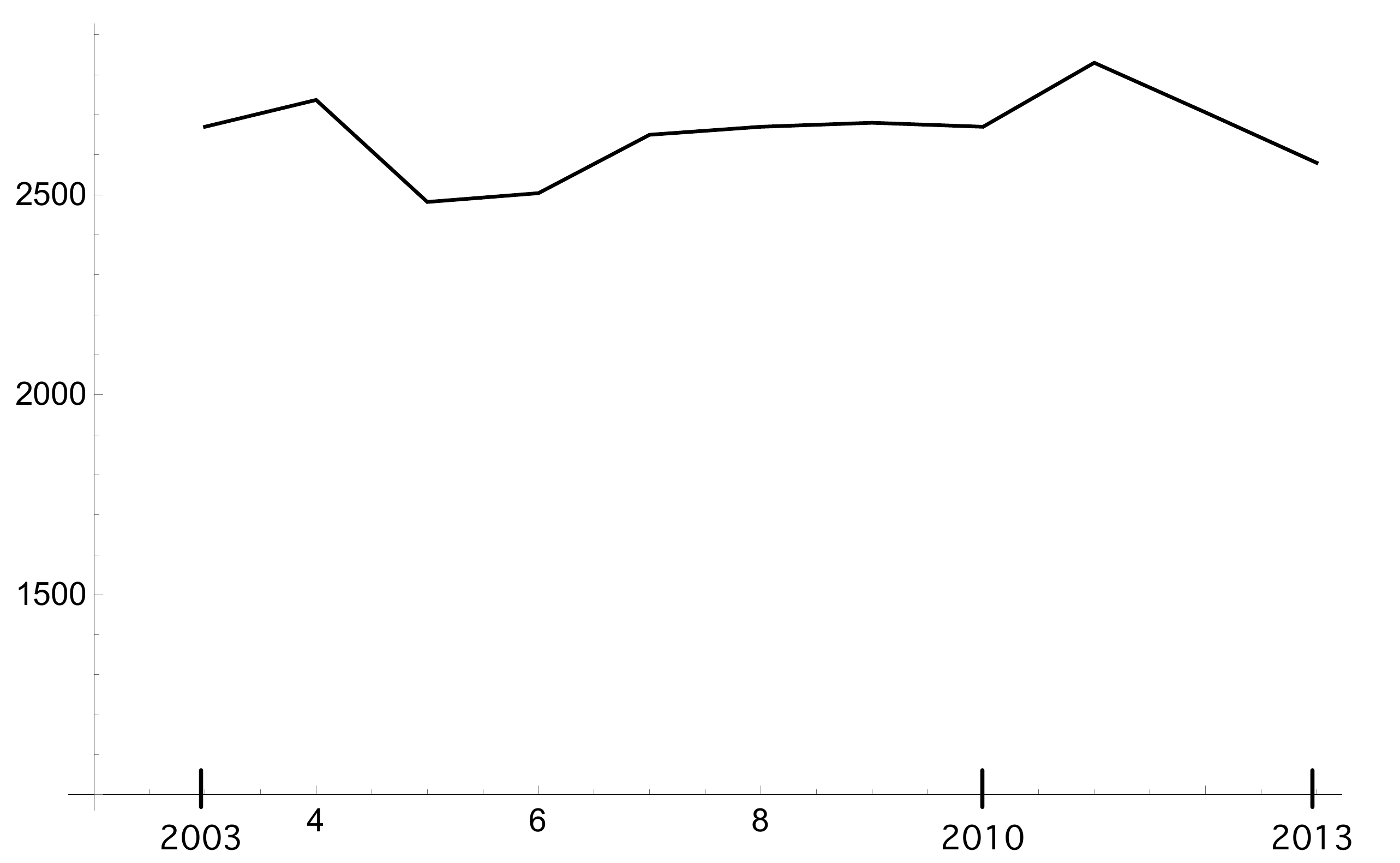}
     \end{center}
  \caption
  {\footnotesize
The number of papers on supersymmetry published in 2003-2013. 
}
\label{fig:pgs4}
\end{figure}

\begin{figure}[h]
  \begin{center}
  \includegraphics[width=3.5in]{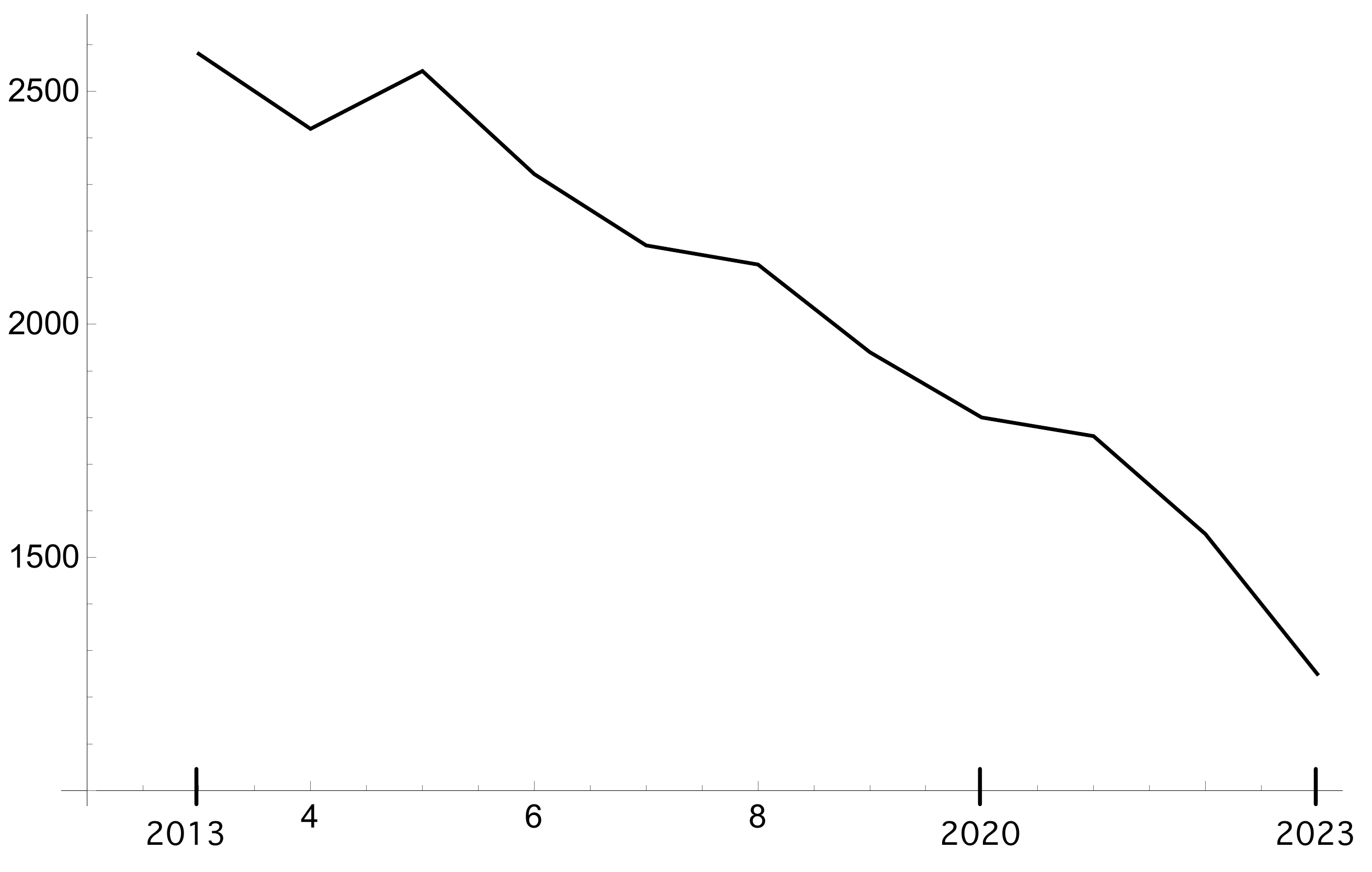}
     \end{center}
  \caption
    {\small
The number of papers on supersymmetry published in 2013-2023.}
\label{fig:pgs5}
\end{figure}

The subsequent explosive proliferation of publications devoted to supersymmetry is presented 
in Figs. \ref{fig:pgs_1}-\ref{fig:pgs5}.  The growth continued until 1987 and then, after a brief decline, resumed in 1993 reaching the absolute maximum 
around 1998. For the subsequent 14 years (until 2012)  the rate of supersymmetry-related papers stabilized roughly at the level of 2700 per year.
Since 2012 we observe a continuous decline which exhibits  a tendency of acceleration in the last few years. The current rate is around 40\% of that at the maximum.
\clearpage
\clearpage

Some features of the plots  in Figs. \ref{fig:pgs_1}-\ref{fig:pgs5}
deserve comments. The steep rise starting in 1982 is apparently correlated with
the appearance of two papers written by E. Witten   \cite{82EDW},
which caused a strong resonance in the community. In 1987 
we see the beginning of a moderate  fall which lasted until  1992.
Presumably, it can be explained by the release of the famous work by
work of P. Candelas {\em et
al.}   \cite{candel} which diverted an active part of the community
to string theory. The papers resulting from this realignment of interests were classified 
 under the topic 
 ``strings'' in the  iNSPIRE-hep data base.
 
The downturn tendency was  overcome by the end of 1992. A quite rapid  and continuous growth
in the publication rate  becomes evident in 1994, after
the breakthrough work of N. Seiberg and E. Witten   \cite{SeWi},
 which appeared in July of 1994. 
For four years fascinating features of supersymmetric gauge
dynamics at strong coupling   \cite{SeWi} captured the imagination of many theorists who seemed to had left the
field. The peak in 1998 is the absolute maximum in the number of publications
(at the level of 2900 publications/year). In 1998/99 we observe a slight decline 
with the subsequent stabilization at the level of 2700. An approximate plateau lasted 
up to 2011. Then, a continuous decline started (Fig. \!\!\ref{fig:pgs5}), with a tendency to acceleration in the last few yeas.
To my mind, the reason lies in the negative results in the searches for superpartners at LHC.
For decades young theorists were investing their efforts to the fullest into phenomenology-based SUSY,
improving MSSM and adding extra features which gave birth to NMSSM, NNMSSM and so on.
Now, being deprived of any hints from nature, they got disheartened. 

 One can say that the phenomenological branch 
of SUSY has almost dried out at the time being (See Fig. \ref{001}, p. \pageref{001}). I do not think this statement to be an exaggeration.
We will have to wait for the experimental discovery of 
supersymmetry (or perhaps some clear-cut hints from nature) to revive the phenomenological studies of SUSY.
The other branch in Fig. \ref{001}, p. \pageref{007}  -- SUSY at strong coupling -- is alive and growing.

The iNSPIRE-hep date base tells us that  the overall number of publication on  supersymmetry during the 52 years that elapsed since its inception
is 58,600. Supergravity adds 16,600.

\newpage

\begin{center}

{\bf\large  FOREWORD TO THE {\color{red} FIRST} EDITION OF\\[1mm]
 ``THE SUPERSYMMETRIC WORLD''
}

\vspace{2mm}
{\large G. L. KANE}$^a$ and  {\large M.  SHIFMAN}$^b$

\vspace{2mm}
$^a$ {\small \em  Randall Laboratory, University of Michigan,  Ann Arbor, MI
48109}

{$^b$\small \em Theoretical Physics Institute, University of Minnesota,
Minneapolis, MN 55455}

{\tt shifman@umn.edu} 
\end{center}

\begin{quote}{\footnotesize
	\hspace{0.7cm} ``... One of the biggest adventures of all is the 
search for supersymmetry. Supersymmetry is the framework in which
theoretical physicists have  sought  to answer some of the questions
left open by the Standard Model of particle physics. \\
\hspace*{0.7cm} Supersymmetry, if it holds in nature, is part of the
quantum  structure of space and time. In everyday life, we measure
space and time by numbers, ``It  is now three o'clock, the elevation is
two hundred meters above see level,'' and so on. Numbers are 	classical
concepts, known to humans since long  before  Quantum Mechanics
was developed in the early twentieth century. The discovery of
Quantum Mechanics changed our understanding of  almost everything
in physics, but our basic way of thinking about space and time has not
yet been affected.\\
\hspace*{0.7cm}  Showing that nature is supersymmetric would change
that, by  revealing a quantum dimension of space and time, not
measurable by ordinary  numbers. .... Discovery of 	supersymmetry would be one of the real milestones in  physics''

\hspace{9cm}  E. Witten   \cite{1} }
\end{quote}

The history of supersymmetry is exceptional.  All other major 
conceptual
developments in physics and science have occurred because
scientists were trying to understand or study some established aspect of 
nature,
or to solve some puzzle arising from data.  
The discovery of supersymmetry in the early 1970's, an invariance of 
the
theory under interchange of fermions and bosons, was 
a purely intellectual achievement, driven by the logic of 
theoretical development rather than by the pressure of  existing 
data.  
Thirty years elapsed from the time of discovery,
immense theoretical effort was invested in this field, over 30,000 papers
published. However, none of them can claim to report the experimental
discovery of supersymmetry (although there are some hints, of which 
we will
say more later). In this respect the phenomenon is 
rather unprecedented in the  history of physics.
 Einstein's general relativity, the closest possible analogy
one can give, was experimentally confirmed within several years after 
its
creation. Only in one or two occasions, have theoretical predictions of 
a comparable magnitude had to wait for experimental confirmation that 
long.
For example, the   neutrino had a time lag
of 27 years. 

It would not be an exaggeration to say that today supersymmetry 
dominates theoretical
high energy physics. Many believe that it will play the same
revolutionary role in the physics of the 21-st century as special and
general  relativity  did in the physics of the 20-th century. 
This belief is based on aesthetical appeal, on indirect evidence,
 and on the fact that no theoretical 
alternative is in sight.

The discovery of supersymmetry presents a dramatic story
dating back to the late 1960's and early '70's. 
For young people who entered high energy physics in the 1990's this 
is ancient history.  Memories fade away as live participants of
these events approach the retirement age; some of them 
have already retired and some, unfortunately, left this world.
Collecting live testimonies of the pioneers,
and preserving them for the future, seems timely given the impact
supersymmetry has already produced on the  development
of particle  physics. 
 Having said that, we note that this book did not appear
 as 
a result of a conscious project.  Both editors had
collected some materials for other activities  \cite{2,3} 
and became aware of the
other's interest and materials. Many people have been interested in how
supersymmetry originated ---the question often is asked in informal
conversations---and how it can be such an active field even
before direct experimental confirmation.  We finally decided to combine
materials, invite further ones, and edit this volume that makes available
a significant amount of information about the origins of this
intellectually exciting area. Most of it is in the words of the original
participants.

In the historical explorations of 
scientific discoveries (especially, theoretical)
it is always very difficult to draw a ``red line'' marking the true
beginning, which would separate ``before'' and ``after.'' 
Almost always there  exists a chain
of works which   interpolates,
 more or less continuously,
between the distant past and the present.  Supersymmetry is no 
exception, the more so because it has multiple roots.
It was observed as a world-sheet two-dimensional 
symmetry\footnote{The realization that
the very same string theories gave rise to supersymmetry in the
target space came much later.}
in string theory  around 1970; at approximately the same time
Golfand and Likhtman found the superextension of the
Poincar\'e algebra and constructed the first four-dimensional
field theory with supersymmetry, (massive) quantum 
electrodynamics of spinors and scalars. Within a year 
Volkov and collaborators (independently) suggested
nonlinear realizations of supersymmtery
and started the foundations of supergravity. Using the
terminology
of  the string practitioners one can say that
 the first supersymmetry revolution
occurred in 1970-71 as the idea originated.\footnote{According to the
Marxist teaching, it would be more appropriate in this case
to speak of a pre-revolutionary situation.
The distinction is too subtle, however, to be discussed in this article.}  The
second supersymmetry revolution came with the work of Wess and Zumino in
1973. Their discovery opened to the rest of the community the gates to the
Superworld. The work on supersymmetry was tightly woven in the fabrique
of  contemporary theoretical physics. During the first few years of its
development, there was essentially no interest in whether or how
supersymmetry might be  relevant
to understanding nature and the traditional goals of physics.  It was 
``a solution in search of a problem.''
Starting in the early 1980's, people began
to realize that supersymmetry might indeed solve some basic problems
of our world.   This  time may be characterized as  the
third supersymmetry revolution.

So, how far in the past one should go and where one should stop
in the book devoted to the beginnings?

The above questions hardly have unambiguous answers.
We decided to start from Ramond, Neveu, Schwarz, Gervais, and
Sakita whose memoirs are collected in the chapter entitled {\em The
Predecessors}, which opens the book. The work of these authors can be 
viewed as precursive to the discovery of supersymmetry in 
four dimensions. It paved the way to Wess and Zumino.

The central in the first part of the book is Chapter 2
presenting  {\em The Discovery}. It contains recollections of
Likhtman, Volkov, Akulov, Koretz-Golfand (Yuri Golfand's widow)
 and the 1999
Distinguished Technion Lecture of Prof. J. Wess, in which 
the basic stages of the
theoretical construction are outlined.\footnote{Unfortunately, our
(probably, awkward) attempts to convince 
  Prof. B. Zumino
in the usefulness of this book failed---we were unable to obtain his
contribution.}  
Chapter 3 is devoted to the advent of supergravity.
The fourth chapter is entitled
{\em The Pioneers}. The definition of pioneers
(i.e. those who made crucial  contributions
at the earliest stage) is  quite ambiguous,  as is the upper cut
off in time which we   
set, {\em the  summer of 1976}. By that time 
no more than a few dozen of original papers on
supersymmetry had been published. 

The selection of the contributors
 was a  difficult task. We were unable to give  floor
to some theorists who were instrumental at the early stages
(e.g. R. Arnowitt, L. Brink, R. Delbourgo, P.G.O. Freund, D.R.T. Jones,
 J.T.~$\mbox{\L}$opusza\'nski,
P. Nath, Y. Ne'eman,  V.I. Ogievetsky, A. Salam, E. Sokat\-chev,  B. de Wit).
Some 
are represented in other chapters (e.g. S. Ferrara
whose 1994 Dirac Lecture is being published   
in Chap. 3.) Others are beyond reach. 
This refers to Abdus
Salam and Victor Ogievetsky.
The latter, by the way, wrote (together with L. Mezincescu)  the first
comprehensive review on supersymmetry which was published in
1975 \cite{ogi}. Even now it remains an
 excellent introduction to the subject, in spite of the 25 years that 
have elapsed.

 The question of where to draw the
line tortured us, and we bring our apologies to all the pioneers who
``fell through the cracks.'' 

The second part of the book is  an attempt to present a historical 
perspective
on the development of the subject. 
This  task obviously belongs to  the professional historians of science;
the most far-sighted of them will  undoubtedly turn their attention to
supersymmetry  soon. For the time being, however,  to the best of our
knowledge, there are no professional investigations on the issue.
There was available a treatise
written by Rosanne Di Stefano in 1988 for a conference proceedings
which were never published. This is a very thorough and insightful 
review.
On the factual side it goes far beyond any other material
on the history of supersymmetry one can find in the literature.
There are some omissions, mostly  regarding the Soviet contributors,
which are naturally explained by the isolation of the Soviet community
before the demise of the USSR
and relative inaccessability of several key papers written in Russian.
The Yuri Golfand Memorial Volume  \cite{3}
which contains the English translation of an important
 paper by  Golfand and
Likhtman  \cite{gl} as well as a wealth of other relevant materials,
fills the gap. In addition, Springer-Verlag
has recently published  Memorial Volumes in honor of Dmitry 
Volkov  \cite{dv} and Victor
Ogievetsky,  \cite{ogim} which acquaint the interested reader with their 
roles
to a much fuller extent than previously. 
 
The coverage of certain physics
issues in Di Stefano's essay required comment; in
a few
cases we added explanatory footnotes. Di Stefano's essay is preceded
by a relatively short article written by the late Prof. Marinov.
It is entitled ``Revealing the Path to the Superworld'' and was
originally  intended for
the Golfand Volume. This article presents ``a bird's eye view''
on the area.
On the factual side it is much less comprehensive than Di Stefano's, but it
carries a distinctive flavor of the testimony of an eye witness.
Moreover, it reveals the mathematical roots of the discovery,
an issue which is only marginally touched in Di Stefano's essay.

We are certainly not professional  historians of science; still
we undertook a little investigation of our own.
Often students ask where the name ``supersymmetry'' came from?
It seems that it was coined in the paper by Salam and Strathdee  \cite{SS}
where these authors constructed supersymmetric Yang-Mills theory.
This paper was received by the editorial office on June 6, 1974,
exactly eight months after that of Wess and Zumino.
Super-symmetry (with a hyphen) is in the title,
while in the body of the paper Salam and Strathdee
use both, the old version of Wess and Zumino,
 ``super-gauge symmetry,'' and the new one.  An earlier paper of Ferrara
and Zumino  \cite{FZ} (received by the editorial office\footnote{The 
editorial
note says it was received  on May 27,
197{\bf 3}. This is certainly a misprint; otherwise, the event would be 
acausal.}
 on May 27,
1974) where the same problem of  super-Yang-Mills  was addressed, 
mentions
only supergauge invariance and supergauge transformations. 

\vspace{0.2cm}

\begin{center}
$\star$ $\star$ $\star$  
\end{center}
Supersymmetry is nearly thirty years old.
It seems that now  we are approaching  the 
fourth supersymmetry
revolution which will demonstrate its relevance to
 nature.  Although not numerous, we do have hints
that this is the case. They are: 
(a) supersymmetry provides a way to
understand how the electroweak SU(2)$\times$U(1) symmetry is 
broken, so
long as the top quark came out heavy (which it did),
(b)  gauge
couplings unify rather accurately when superpartners are included in 
the
loops,\footnote{An alternative way to say this is to say that the value of 
the
weak mixing angle at the weak scale can be calculated accurately if one 
sets it
to the value predicted by a unified theory at the unification scale} (c) 
the 
Higgs boson is predicted to be light  (LEP gives $M_H <200$ GeV),
and (d) the lack of any deviations from Standard
Model predictions in the precision data at LEP and 
in other experiments
is consistent with supersymmetry (it was anticipated that these
 deviations 
would be invisible).

Certainly,  at the moment the indications are not conclusive.
However inconclusive, they are the source of hope
and enthusiasm for 
phenomenologically oriented theorists and experimentalists
who would like to
keep high-energy physics in the realm of empirical science.

Another aspect which came to limelight recently
 is the fact that supersymmetry became
instrumental in the solution of highly nontrivial dynamical issues in
strongly coupled non-supersymmetric theories, which defied
solutions for decades.  That of course does not imply that
nature is supersymmetric, but it does add to the interest in
supersymmetry.

Summarizing, in this book
we bring together contributions from many of the key players of the 
early
days of supersymmetry.
We leave its relevance 
 to our world to a future project.
 
\vspace{0.5cm}

\section*{List of participants}

V.~Akulov, P.~Fayet, S.~Ferrara, D.~Freedman, G.-L.~Gervais,
J.~Iliopoulos,  G.L.~Kane,
N.~Koretz-Golfand, E.~Likhtman, M.~Marinov, A.~Neveu,
P.~van~Nieuwenhuizen, L.~O'Raifeartaigh,
P.~Ramond,
B.~Sakita, J.~Schwarz, J.~Wess, M.~Shifman, M.~Sohnius, V.~Soroka, J.~Strathdee,
R. Di Stefano, D.~Volkov, P.~West.

\newpage

{\bf Quotes from the First Edition reviews:}

\vspace{2mm} 

This book [...] collects personal reminiscences of the pioneers and founders of supersymmetry. How many people know why supersymmetry was first introduced in particle physics, or how superstrings were invented (by Ramond, Neveu and Schwarz) before supersymmetry was even known, or, in a more anectodal vein, how the name developed from the `super-gauge symmetry' of Wess and Zumino to super-symmetry? (with the hyphen) of Salam and Strathdee? This book is excellent reading for all of those who do not know the answers or just want to know more.

\vspace{1mm}

\hspace{3cm} Gianfrancesco Giudice, CERN Courier, 29 August 2001

\begin{center}
******
\end{center} 
I appreciated the personally written contributions the most -- many quite graphic pictures of the past were elucidated and received new shades. There are many interesting ``anecdotes,'' in fact, they are almost countless. It is, however, not a gossiping book, the factual material predominates. [If supersymmetry is found] then the book will become essential reading for the Nobel Committee. Otherwise, it will remain a monument to incredibly long wanderings in search of a mirage in the desert of theoretical physics.
\vspace{1mm} 

\hspace{5.6cm} Claus Montonen,
  University of Helsinki

\vspace{1mm}
\footnotesize 
\hspace{1cm}  {\sl Arkhimedes},  Fysiikan ja matematiikan aikakauslehti (Journal of Physics and 
\hspace{1cm}  Mathematics), 2001 : 6 / 2002 : 1, s. 32 (in Swedish).

English translation in https://www-users.cse.umn.edu/~shifman/Montonen.rev.pdf

\end{document}